\documentstyle[12pt]{article}

\newcommand{\sect}[1]{\setcounter{equation}{0}\section{#1}}

\def\be{\begin{eqnarray}}
\def\ee{\end{eqnarray}}
\def\Z{{\cal Z}}
\def\C{{\cal C}}
\def\L{{\cal L}}

\font\twelvemsa=msam10 scaled 1200

\font\sevenmsa=msam7
\font\fivemsa=msam5
\newfam\msafam
\textfont\msafam=\twelvemsa
\scriptfont\msafam=\sevenmsa
\scriptscriptfont\msafam=\fivemsa
\def\msa{\ifcase\msafam 0\or1\or2\or3\or4\or5\or6\or7\or8\or9\or A\or B\or
C\or D\or E\or F\fi}

\font\twelvemsb=msbm10 scaled 1200

\font\sevenmsb=msbm7
\font\fivemsb=msbm5
\newfam\msbfam
\textfont\msbfam=\twelvemsb
\scriptfont\msbfam=\sevenmsb
\scriptscriptfont\msbfam=\fivemsb
\def\msb{\ifcase\msbfam 0\or1\or2\or3\or4\or5\or6\or7\or8\or9\or A\or B\or
C\or D\or E\or F\fi}

\font\twelveeuf=eufm10 scaled 1200

\font\seveneuf=eufm7
\font\fiveeuf=eufm5
\newfam\euffam
\textfont\euffam=\twelveeuf
\scriptfont\euffam=\seveneuf
\scriptscriptfont\euffam=\fiveeuf
\def\euf{\ifcase\euffam 0\or1\or2\or3\or4\or5\or6\or7\or8\or9\or A\or B\or
C\or D\or E\or F\fi}

\mathchardef\gapprox"3\msa26
\mathchardef\lapprox"3\msa2E

\begin{document}

\title{
Chiral models in
dilaton--Maxwell gravity
}

\author{~Oleg ~V. ~Kechkin and ~Maria ~V. ~Yurova\\ 
\\D\,E\,P\,N\,I
\\Institute of
Nuclear Physics\\Moscow State University
\\Vorobjovy Gory\\119899 Moscow\\Russia}

\maketitle

\begin{abstract}
\noindent
We study symmetry properties of the Einstein--Maxwell theory nonminimaly
coupled to the dilaton field. We consider a static case with pure electric
(magnetic) Maxwell field and show that the resulting system becomes a
nonlinear $\sigma$--model wich possesses a chiral representation.
We construct
the corresponding chiral matrix and establish a representation which is
related to
the pair of Ernst--like potentials. These potentials are used for
separation of the symmetry group into the gauge and nongauge (charging)
sectors. New variables, which linearize the action of charging symmetries,
are also established; a solution generation technique based on the use of
charging symmetries is formulated. This technique is used for generation
of the elecricaly (magneticaly) charged dilatonic fields from the static
General Relativity ones. 
\end{abstract}
\renewcommand{\thepage}{ }
\pagebreak

\renewcommand{\thepage}{\arabic{page}}
\setcounter{page}{1}


\sect{Introduction}
It is well known fact that the Kaluza--Klein, supergravity and
perturbative (super)string theories
provide a large number of the effective four--dimensional gravity models
\cite {kk}, \cite {ss}.
These models describe various interacting scalar, vector and tensor fields
coupled to gravity. In the simplest case of the single scalar and gauge
fields the corresponding action reads
\be
^4S=\int d^4x |g|^{\frac{1}{2}}\left [ -^4R+2\left (\partial\phi\right )^2
-e^{-2\alpha\phi}F^2\right ],
\ee
where $F_{\mu\nu}=\partial_{\mu}A_{\nu}-\partial_{\nu}A_{\mu}$. Here
$\alpha=\sqrt 3$ in the case of the five--dimensional Kaluza--Klein theory 
compactified on a spatial--like circle and $\alpha=1$ for the
naturally (i.e., with zero moduli) truncated low--energy heterotic string 
theory without axion field as well as for the N=2, D=4 supergravity. Other
values of the parameter $\alpha$ can arise in the framework of less
convenient truncations of multidimensional or (super)string theories
compactified to four dimensions, so the study of the theory (1.1) is
interesting  for applications.

In \cite {gk1} it was shown that this theory in the stationary case
becomes a non--linear $\sigma$--model. It was also shown that this
$\sigma$--model possesses a symmetry group which contains a gauge part
for the arbitrary value of $\alpha$. However, only in the Kaluza--Klein
theory case it arises also a sector of the
nongauge symmetries, which consists of the Ehlers--Harrison type
transformations \cite {maison}. The presense of this nongauge and nonlinear
sector is closely related to the appearance of integrable properties
which
arise after the following reduction to two dimensions (i.e. for the
stationary
and axisymmetric fields). In fact, for integrable systems it
exists an
infinite set of the conserving quantities; much of powerful and perfect
methods can be applied to such systems (see \cite {is} for the inverse
scattering transform technique and \cite {backlund} for the Bachlund
transformation).

Below we consider the static case, i.e. the case when all fields do not
depend  on a time and the four--dimensional line element can be parametrized
as
\be
ds^2=g_{\mu\nu}dx^{\mu}dx^{\nu}=fdt^2-f^{-1}h_{ij}dx^idx^j=fdt^2-f^{-1}dl^2.
\ee
Moreover, we will study pure electric or pure
magnetic configurations (i.e. we will suppose that $A_t\neq 0, A_i=0$ or
$A_t=0, A_i\neq 0$). As it was shown in \cite {gk1}, in both cases the
theory (1.1) leads to three--dimensional model which can be described
by the action
\be
^3S=\int d^3x h^{\frac{1}{2}}\left [ -^3R+^3\L\right ],
\ee
where $^3R$ is a Ricci scalar constructed using $h_{ij}$ and
\be
^3\L=2\left (\nabla\phi\right )^2+
\frac{1}{2}f^{-2}\left ( \nabla f\right )^2-f^{-1}e^{\pm 2\alpha\phi}
\left (\nabla w\right )^2.
\ee
In the ``matter'' Lagrangian (1.4) the signs ``$-$'' and ``$+$'' 
are related to the electric and magnetic cases correspondingly;
for the electric case
\be
w=\sqrt 2A_t,
\ee
whereas for the magnetic one
\be
\nabla w=\sqrt 2e^{-2\alpha\phi}f\nabla\times\vec A.
\ee
Thus, $w$ is the electric (magnetic) potential;
so it is a scalar (pseudoscalar) field from the three--dimensional point
of view.

In the next section we construct several representations of the theory
(1.3)--(1.4) to clarify its symmetry properties in the case of
$\alpha\neq 0$. It will
be shown that for the arbitrary value of $\alpha$ one obtains
a chiral model which is
closely related to the one appearing in the framework of
pure (nonstatic) General Relativity \cite{k1}.

In \cite {gk1} it was shown that the complete three--dimensional
theory, corresponding to (1.1) (i.e. the extension of Eq. (1.4) to the case of
$g_{ti}\neq 0$ and both $A_t\neq 0$, $A_i\neq 0$), in the string theory
case 
has
not the nongauge symmetries and does not become integrable in two dimensions.
However, the including of the axion field into the action (1.1) in a
way predicting by the 
heterotic string theory provides a ``right'' correction of the theory
(1.3)--(1.4):
the resulting model possesses a reach symmetry structure in three and
two dimensions \cite {gk2}.


\sect{Gauge and charging symmetries}

First of all, from Eq. (1.4) it follows that the discrete transformation
\be
\phi\rightarrow -\phi
\ee
maps the electric system into the magnetic one and overwise. In fact,
this discrete symmetry is the ``part'' of the continuous electric--magnetic
duality, which exists for the complete (i.e., electric--magnetic)
effective three--dimensional
theory. To establish other symmetries let us introduce new
functional variables $F$, $\Phi$ and $W$ accordingly to the formulae
\be
F=fe^{\mp 2\alpha\phi},\quad
e^{2\Phi}=f^{\pm\alpha}e^{2\phi},\quad
W=\sqrt{\frac{1+\alpha^2}{2}}w.
\ee
In terms of these variables
\be
^3\L&&=\frac{^3L}{1+\alpha^2},\quad{\rm where}
\nonumber\\
^3L&&=2\left (\nabla\Phi\right )^2+
\frac{1}{2}F^{-2}\left (\nabla F\right )^2-2F^{-1}
\left (\nabla W\right )^2.
\ee
One can see that $^3L$ is the sum of two noncoupled Lagrangians; the
uniquely symmetry of the first one $^3L_1=2(\nabla\Phi)^2$ is
\be
\Phi\rightarrow\Phi+\lambda_0,
\ee
where $\lambda_0$ is the arbitrary real parameter. To establish the symmetry
group of the second Lagrangian $^3L_2=
\frac{1}{2}F^{-2}\left (\nabla F\right )^2-2F^{-1}
\left (\nabla W\right )^2$ it is useful to introduce the functions
\be
 E_{\pm}=F^{\frac{1}{2}}\pm W;
\ee
in terms of them
\be
^3L_2=8\frac{\nabla E_+\nabla E_-}
{(E_++E_-)^2}.
\ee
From this form of $^3L_2$ it immediately follows that
\be
E_{\pm}\rightarrow e^{\lambda_1}E_{\pm}\quad {\rm (scaling)}\quad {\rm and}
\quad
E_{\pm}\rightarrow E_{\pm}\pm\lambda_2 \quad {\rm (shift)}
\ee
are the symmetries for the arbitrary real parameters $\lambda_1$ and
$\lambda_2$.
Then, the map
\be
E_{\pm}\rightarrow E_{\pm}^{-1}
\ee
is the important discrete symmetry of Eq. (2.6), this symmetry generalizes
the
corresponding transformation established by Kramer and Neugebauer in
\cite {kn} for the
stationary General Relativity (see also \cite {hk1} for the 
complete, i.e., with the nonzero moduli, heterotic string theory case).
It is easy to
see that Eq. (2.8) maps the
scaling transformation into itself, whereas shift becomes the 
Ehlers--like symmetry \cite {ehlers}
\be
E_{\pm}^{-1}\rightarrow E_{\pm}^{-1}\pm\lambda_3 \quad {\rm (Ehlers)}
\ee
(here $\lambda_3$ is the arbitrary\, real\, parameter).
Thus, we obtain three
one--parametric symmetry transformations for the system (2.6); one can
prove that their generators form an algebra of the group ${\rm SL(2,R)}$.

Let us now introduce the following $2\times 2$ matrix
\be
M=F^{-\frac{1}{2}} 
\left (\begin{array}{crc}
1& \quad & W\\
W &\quad & W^2-F
\\
\end{array}\right ),
\ee
then
\be
^3L_2={\rm Tr}\left ( J_M^2\right ) \quad {\rm with}
\quad J_M=\nabla M\,M^{-1}.
\ee
The matrix $M$ is a symmetric matrix of the signature $+\,-$; its
determinant is equal to $-1$. The general symmetry
transformation preserving these properties is
$M\rightarrow C^TMC$  with ${\rm det}\,C=1$,
so $C\in SL(2,R)$ as it was noted before. One can prove that
the matrix ${\cal M}=e^{\Phi}M$ provides a chiral representation of the
whole $^3L$. Actually, in view of Eqs. (2.3) and (2.11) one immediately
obtains
that $^3L={\rm Tr}J_{\cal M}^2$, where $J_{\cal M}=\nabla {\cal M}\,
{\cal M}^{-1}$. Then the general symmetry transformation reads
\be
{\cal M}\rightarrow {\cal C}^T{\cal M}{\cal C}
\ee
with ${\cal C}\in GL(2,R)$. Here the additional $U(1)$ transformation
exactly corresponds to the one of Eq. (2.4); this $U(1)$ transformation moves
the $\Phi$--asymptotics. It is not difficult to prove that the
parametrization of ${\cal C}$ in terms of the before introduced
parameters $\lambda_{\mu},\,\,\mu=0,...,3$ reads:
\be
{\cal C}=e^{\frac{\lambda_0-\lambda_1}{2}} 
\left (\begin{array}{crc}
1& \quad & \lambda_2\\
\lambda_3 &\quad & e^{\lambda_1}+\lambda_2\lambda_3
\\
\end{array}\right ).
\ee

Let us now establish symmetries which preserve asymptotical flatness
property of the field configurations, i.e., the charging symmetries.
Thus, we consider fields with the spatial asymptotics
$f_{\infty}=1$ and $\phi_{\infty}=w_{\infty}=0$,
i.e., we suppose that $\Phi_{\infty}=0$ and $(E_{\pm})_{\infty}=1$.
It is easy to see that
for the corresponding transformations $\lambda_0=0$.
Then, after some algebra one
obtains that the remaining $SL(2,R)$ transformations do not change
asymptotics
only if $\lambda_2=\lambda_3\equiv\lambda$ and $e^{\lambda_1}=1-\lambda^2$.
In this case
\be
{\cal C}=\frac{1}{\sqrt{1-\lambda^2}} 
\left (\begin{array}{crc}
1& \quad & \lambda\\
\lambda &\quad & 1
\\
\end{array}\right ).
\ee
The
corresponding transformations of the Ernst potentials read:
\be
E_{\pm}\rightarrow\frac{E_{\pm}\pm\lambda}{1\pm\lambda E_{\pm}}.
\ee

From Eq. (2.14) it follows that the charging symmetry transformation
is the boost
with the velocity parameter $\zeta /2= {\rm Arcth}\,\lambda$.
It is possible to linearize it
by the choose of new appropriate field variables. Using the evident
analogy of the Lagrangian (2.6) with the one of stationary
General Relativity (see \cite {k1} for the details) it is natural to take
\be
\Z_{\pm}=\frac{1-E_{\pm}}{1+E_{\pm}};
\ee
then 
\be
\Z_{\pm}\rightarrow e^{\mp\zeta}\Z_{\pm}.
\ee
These formulae together with the condition $\Phi={\rm inv}$
completely define the action of charging symmetry on
the potentials. All the remaining symmetries move field asymptotics
and form a gauge sector of the complete symmetry group.
One can also consider the action of charging symmetry
transformation on
the charges, which we define accordingly the expansions $f\rightarrow
1-2M_{gr}/r,\,\phi\rightarrow D/r,\,A_t\rightarrow Q_e/r$ (we consider the
electric case for definiteness). Then, as it can easily be checked, the
combination $D-\alpha M_{gr}$ remains invariant, whereas the quantities
$M_{gr}+\alpha D\mp\sqrt {1+\alpha^2}Q_e$ transform exactly as the
potentials $\Z_{\pm}$.

\sect{Solution generation technique}
The variables $\Z_{\pm}$ together with $\Phi$
form the most natural set of potentials for the asymptoticaly flat field
configurations.
The procedure for generation of the asymptoticaly flat
solutions from the known ones becomes trivial in terms of these
variables. For example, if one starts from the pure General Relativity
with the static line element (1.2), and performs the charging symmetry
transformation, one obtains the following solution:
\be
ds^2=\frac{f}{{\rm H}^{\frac{2}{1+\alpha^2}}}dt^2-
\frac{{\rm H}^{\frac{2}{1+\alpha^2}}}{f}dl^2,
\nonumber
\ee
\be
\phi=
\frac{\alpha}{1+\alpha^2}{\rm ln}
\,{\rm H},
\quad
A_t=
\frac{\sinh\zeta}{2\sqrt{1+\alpha^2}}\,\frac{1-f}{\rm H},
\ee
where ${\rm H}=[1+f+(1-f)\cosh\zeta]/2$.
Thus, the transformation preserving asymptotical flatness (we suppose
that $f_{\infty}=1$) generates the nontrivial electric and dilaton
potentials (for obtaining of the magnetic solution one must apply
the map (2.1) and replace $A_t$ to $\vec A$, see Eq. (3.10) below).
In particular, from the massive solutions one obtains the
electricaly (magneticaly) charged ones; so these transformations
are actually charging transformations \cite {kin} (see also
\cite {hk1} for the heterotic string theory case). 

Now let us consider the action of the full group of symmetry 
transformations on the set
of divergent--free currents uniquely related to the problem. First of
all, the equations of motion can be written as
\be
\nabla J_0=\nabla J_1=\nabla J_2=0,
\ee
where
\be
J_0=\nabla\Phi, \quad J_1&&=F^{-1}\nabla \left ( F-W^2\right ),
\nonumber\\
J_2&&=F^{-1}\nabla W.
\ee
It is easy to prove that these currents together with
\be
J_3=F^{-1}\left [ \left ( F+W^2\right )\nabla W-W\nabla F\right ]
\ee
form the matrix current $J_{\cal M}$. Actually, the
straightforward calculation leads to
\be
J_{\cal M}=
\left (\begin{array}{crc}
J_0-J_1/2 & \quad & -J_2\\
J_3 &\quad & J_0+J_1/2
\\
\end{array}\right );
\ee
so from the equation $\nabla J_{\cal M}=0$ it follows that
$\nabla J_3=0$. In fact, this last equation is the algebraic consequence
of Eqs. (3.2); however the current $J_3$ is actually independent on the
currents $J_0, J_1$ and $J_2$. These four algebraicaly independent
currents linearly transform under the group of $GL(2,R)$ symmetries,
because from Eq. (2.12) it follows that
\be
J_{\cal M}\rightarrow {\cal C}^T
J_{\cal M}\left ( {\cal C}^{T}\right )^{-1}.
\ee
Now let us consider the charging symmetry transformation with the
matrix $\C$ written in $\zeta$--terms. Then, after some algebraical
analysis one leads to the following result: two current combinations
\be
J_{\pm}=J_2+J_3 \pm J_1
\ee
transform exactly as $\Z_{\pm}$ (see Eq. (2.17)),
\be
J_{\pm}\rightarrow e^{\mp\zeta}J_{\pm},
\ee
whereas $J_0$ and $J_2-J_3$ ones remain invariant.
Let us also note that
\be
\nabla\times\vec A=\frac{J_2}{\sqrt{1+\alpha^2}}.
\ee
(see Eqs. (1.6), (2.2) and (3.3)), so the calculation of $\vec A$
for transformed magnetic solution is equivalent to the calculation
of the transformed current $J_2$.
Then, for the magnetic variant of the above formulated generation
procedure (see Eqs. (3.1) and remember the map (2.1)) one obtains that
\be
\vec A=-
\frac{\sinh\zeta}{2\sqrt{1+\alpha^2}}\vec\Lambda
\ee
Here $\nabla\times\vec\Lambda=f^{-1}\nabla f$ is the single nonvanishing
current of the original static solution of General Relativity.
For example, if one starts from the Schwarzshild solution,
i.e., if one takes
$f=1-2m/r$ and $dl^2=dr^2+r(r-2m)(d\theta^2+\sin^2\theta\,d\varphi^2)$,
then the only nonzero $\vec\Lambda$--component is $\Lambda_{\varphi}=
-2m\cos\theta$. This result is more than natural in the framework of the
magneticaly charged dilatonic black hole physics, as well as all the
formulae (3.1), see \cite {dg}.

\sect{Discussion}
In this work we have established the full symmetry group of the static
electric and magnetic sectors of the four--dimensional dilaton--Maxwell
gravity. From this symmetry group we have extracted a subgroup which
preserves an asymptotical flatness property and have established the action
of this subgroup on the potentials and currents. We have found the
special potential and current combinations ($\Z_{\pm}$ and $J_{\pm}$)
which extremely simplify the action of charging symmetries and, moreover,
transform in the same way, see Eqs. (2.17) and (3.8). It was also
shown that the
remaining potential and current degrees of freedom form
invariants of the 
charging symmetry transformations. Using the developed formalism we
have constructed the charging symmetry invariant extension of the static
Einstein fields to the static electric and magnetic dilaton--Maxwell
gravity. The established formulae (3.1) and (3.10) can be used, for
example, in the black--hole physics after the substitution of the concrete
values of $f$ and $h_{ij}$.

The effective theory (1.3)--(1.4), as a model possessing a
chiral representation,
becomes completely integrable after the following reduction
to two dimensions (for example, in the axisymmetric case). This means, in
particular, an appearance of the infinite--dimensional symmetry group
and the infinite number of the divergent--free currents. This symmetry group,
which is the analogy of the General relativity Geroch group \cite {geroch},
can be obtained
using the spatial localization of the global transformation (2.12) in a
way similar to one established for the principal chiral fields
in \cite {pcf}. We hope to
perform the corresponding analysis in the forthcoming publication.


\bigskip

\hfil{\large\bf Acknowledgements}\hfil

\hskip 3mm

This work was supported by RFBR grant ${\rm N^{0}} 
\,\, 00\,02\,17135$. 

\medskip

\noindent


\end{document}